\documentclass[poms,nonblindrev]{preprint} 

\OneAndAHalfSpacedXI 

\usepackage{color}
\definecolor{EditBlue}{RGB}{0,128,255}

\usepackage{url}
\usepackage{graphicx}
\usepackage{rotating}
\usepackage{tikz}
\usepackage{lscape}
\usepackage{caption}

\usepackage{subfigure, epsfig}
\usepackage{natbib}
\usepackage{tabto}
\usepackage{multicol}
 \bibpunct[, ]{(}{)}{,}{a}{}{,}%
 \def\bibfont{\small}%
\TheoremsNumberedThrough     
\ECRepeatTheorems

\EquationsNumberedThrough    

\usepackage{booktabs}
\usepackage[utf8]{inputenc}
\begin{document}


\RUNAUTHOR{Dimas, Konrad, Maass and Trapp}


\TITLE{Operations Research and Analytics to combat human trafficking: A systematic review of academic literature}

\ARTICLEAUTHORS{%
\AUTHOR{Geri L. Dimas}
\AFF{Data Science Program, Worcester Polytechnic Institute} 
\AUTHOR{Renata A. Konrad}
\AFF{Business School, Worcester Polytechnic Institute} 
\AUTHOR{Kayse Lee Maass}
\AFF{Mechanical and Industrial Engineering Department, Northeastern University} 
\AUTHOR{Andrew C. Trapp}
\AFF{Business School, Worcester Polytechnic Institute}
\AFF{Data Science Program, Worcester Polytechnic Institute}
} 
\ABSTRACT{%
Human trafficking is a widespread and compound social, economic, and human rights issue occurring in every region of the world. While there have been an increasing number of anti-human trafficking works from the Operations Research and Analytics domains in recent years, no systematic review of this literature currently exists. We fill this gap by providing a systematic literature review that identifies and classifies the body of Operations Research and Analytics research related to the anti-human trafficking domain, thereby illustrating the collective impact of the field to date. We classify 142 studies to identify current trends in methodologies, theoretical approaches, data sources, trafficking contexts, target regions, victim-survivor demographics, and focus within the well-established 4Ps principles. Using these findings, we discuss the extent to which the current literature aligns with the global demographics of human trafficking and identify existing research gaps to propose an agenda for Operations Research and Analytics researchers.
}%

\KEYWORDS{human trafficking, operations research, analytics, survey, literature review} 

\maketitle

%


\section{Introduction}
\label{Intro_1}
Human trafficking (HT) involves the commercial exchange and exploitation of individuals for monetary or other gain using force, fraud, or coercion \citep{omt19} and is a widespread social, economic, and human rights issue. While the trafficking of individuals is a centuries-old phenomenon, over the past two decades there has been growing public and research awareness, in part with the ratification of the 2000 Palermo Protocol to Prevent, Suppress, and Punish Trafficking in Persons \citep{ohc00}. Although precise figures are elusive, the Global Estimates of Modern Slavery Report estimates that HT impacts 25M individuals and annually generates more than 150 billion USD in illicit gains globally \citep{int17,int14}. HT is broadly classified as labor and sex trafficking; while all trafficking features exploitation, the actions and means by which HT occurs may differ \citep{gal20}. Labor trafficking takes place in a wide variety of sectors, including the agriculture, domestic work, construction, fishing, food service, and beauty industries. Sex trafficking is a part of the broader commercial sex industry, occurring in industries such as escort services, brothels, and pornography.

Because the scope of HT activity is vast and there are diverse ways in which individuals are exploited \citep{pol17}, context is critical, and effectively addressing HT increasingly requires efforts from multiple disciplines, including interdisciplinary collaborations. For example, HT interventions include approaches from multiple sectors and disciplines such as social work \citep{ama20,oke18,pot10}, healthcare \citep{fra20,hem16}, criminal justice \citep{mcc16,far14,far09}, and economics \citep{de14,mah10}; each domain brings unique perspectives and methods to understand and address HT.

Owing to the breadth of domains that contribute to anti-HT research, a wealth of literature exists that has been well-documented in surveys over the years \citep{mca21,fra20,wen20,mah19,sza18,hem16,sch06}. Existing reviews focus on social science, healthcare, and law enforcement approaches; whereas OR and Analytics have much to offer \citep{kon17}, no review exists for the emerging landscapes of Operations Research (OR) and Analytics as applied to HT.

The present study identifies and classifies the existing OR and Analytics literature related to anti-HT operations and proposes an agenda for future research in this field, filling a gap in the current literature. This work focuses on the four broad principles of anti-trafficking: prevention, protection, prosecution, and partnership (4Ps)~\citep{omc2016}, extending their definition in relation to the OR and Analytics fields. We examine the following research questions:

\tabto{2em} (i) What aspects of HT are being studied by OR and Analytics researchers?
\tabto{2em}(ii) What OR and Analytics methods are being applied in the anti-HT domain? 
\tabto{2em}(iii) What are the existing research gaps associated with (i) and (ii)?
 
We organize the remainder of our study as follows. In Section \ref{Method_2} we define the scope of this review, and in Section \ref{Data_3} we define the data features for analysis. In Section \ref{Implications_4} we discuss the implications of the survey and, based on the observed gaps, suggest areas for future work. We conclude our study in Section \ref{Conclusions}. 

\section{Method of Collection and Categorization}
\label{Method_2}
\begin{figure}[!ht]
		\centering
	\caption{Process of Data Collection.}
	\vspace{0 in}
 \includegraphics[trim = 0 0 0 0, scale=1]{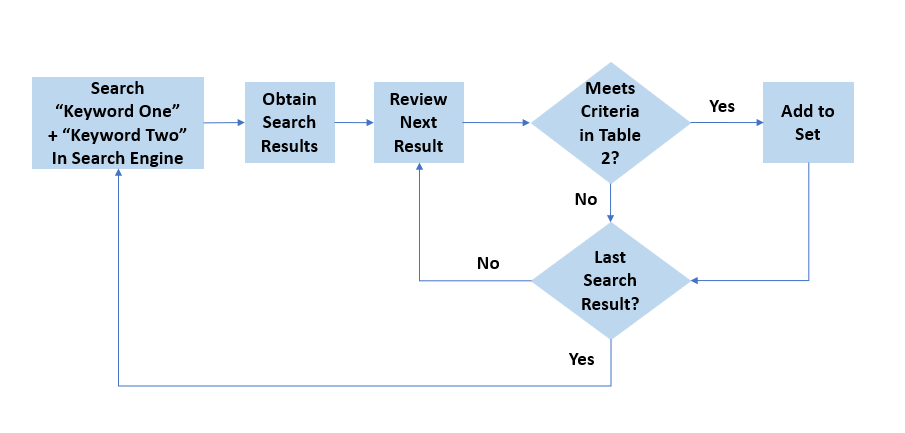}		\label{fig:Figure1}
\end{figure}

We conducted a systematic literature review inventorying studies to answer the three research questions outlined in \ref{Intro_1}.
The methodology used for this systematic review was guided by the Preferred Reporting Items for Systematic Reviews and Meta-Analyses (PRISMA) \citep{moher09}. The collection process (Fig \ref{fig:Figure1}) was based on keyword searches that generated a set of research studies for analysis. Two sets of search words were defined using the combined knowledge of the authors on HT terminology and OR and Analytics methods. These keywords were used in a procedure to identify and select studies that met a set of pre-defined criteria. The first set of keywords reflects terms related to HT, while the second reflects common methods in the OR and Analytics fields (see Table
\ref{table1}). The search and selection of studies was performed by the lead author (G.L.D.), and any uncertainty regarding a study's inclusion was resolved through discussion with the coauthors.

Each search query followed the format: “Keyword One” + “Keyword Two” (such as “Debt Bondage” AND “Integer Programming”), each keyword pair was applied across three bibliographic databases: Scopus, Web of Science, and Google Scholar. The database search was conducted from June 2021 through March 2022. The search results were truncated to studies available through the end of 2021 to provide a comparable year-over-year basis for the research landscape.The sum of these two searches resulted in a total of 449,407 studies for potential inclusion. After the keyword search identification process, a two-step selection process was followed (see Fig \ref{fig:Figure2}). An initial screening process was conducted that evaluated the search results returned for each query, where titles and abstracts were screened and added to the set based on the criteria outlined in Table \ref{table2}. After the initial screening process, the set included 230 unique studies for review. A more in-depth review using the eligibility requirements checklist (Table \ref{table2}) was followed in step two for each of the 230 studies.

\begin{table}[!ht]
\caption{\bf{Keywords Used in Search Process.}}
\centering
		{\def\arraystretch{1.5}  
			\scriptsize{} 
\begin{tabular*}{1\textwidth}{@{\extracolsep{\fill}}ll}
\toprule
\multicolumn{2}{c}{Keyword One: Human Trafficking Related}\\
\midrule
Child Labor / Child Labour & Debt Bondage            \\
Domestic Servitude & Forced Labor/ Forced Labour\\
Human Trafficking  & Labor Trafficking / Labour Trafficking \\
Modern Slavery & Sex Trafficking \\
Trafficking in Persons   \\
\midrule
\multicolumn{2}{c}{Keyword Two: Methodology Related}\\
\midrule
Clustering/Classification & Data Envelopment Analysis \\
Data Science & Game Theory \\
Graph Theory/Construction & Information Extraction \\
Integer Programming & Machine Learning \\
Natural Language Processing  & Network Interdiction/Flow\\
Operations Research & Queueing/Queueing Theory\\
Resource Allocation  & Simulation\\
Simulation & Supervised/Unsupervised Learning\\
Supply Chain & Web Crawling\\
\bottomrule
\end{tabular*}
\label{table1}
}
\end{table}

\begin{table}[!ht]
\caption{Eligibility Requirements Checklist.}
\centering
		{\def\arraystretch{1.5}  
			\scriptsize 
\begin{tabular*}{1\textwidth}{@{\extracolsep{\fill}}lll}
\toprule
Requirements Checklist\\
\midrule
1. Main contribution or focus fell into one of the following three themes:\\

\begin{minipage}[t] {0.9\textwidth} 
\begin{itemize}
 \setlength\itemsep{0.5em}
    \item \textit{Methodological} Operations Research orientation
    \item \textit{Methodological} Analytics orientation (Data science, or other applied analytics)
    \item \textit{Position / Thought} pieces in Operations Research or Analytics
\end{itemize}
\end{minipage}
\vspace{1mm}
\\2. Main application or case study was on anti-HT efforts
\\ 3. Only studies, articles, theses and dissertations were kept (e.g., no books, workshops or government reports) 
\\4. If multiple versions of a study exist (such as a conference paper followed by a peer-reviewed journal article) only the most\\
\hspace{1em} recent, comprehensive version was kept 
\\5. If the study appeared as a section in one study but was further developed into a full study, only the full study was kept \\
\bottomrule
\end{tabular*}
\label{table2}
}
\end{table}

First, only studies that fell into one of the three themes: \emph{Operations Research} methodologies, \emph{Analytical} methodologies, or \emph{Position / Thought} pieces related to Operations Research and Analytics were included. Second, only studies whose primary application area was HT and was written completely in English text were included. Third, book chapters, workshops, and government reports were excluded. Finally, if multiple studies related to a single work were found, only the most complete version were included. Peer-reviewed studies, dissertations, and pre-print studies were included to produce a full and comprehensive review of the current research landscape. The full article screening process resulted in a total of 142 studies included in the set for final review.

As with any search process based upon a predefined set of keywords and checklist requirements, the 142 identified studies may not be exhaustive in scope. However, given our collective experience in researching at the intersection of anti-HT and OR and Analytics, we believe the generated set of studies is representative of current literature at this confluence. A repository containing our classification data can be found publicly at {\url{https://github.com/gldimas/Dimas-et.-al-2022_Human-Trafficking-Literature-Review}.

\begin{figure}[!ht]
		\centering
	\caption{Overview of Selection Process.}
	\vspace{0 in}
 \includegraphics[trim = 0 0 0 0, scale=1]{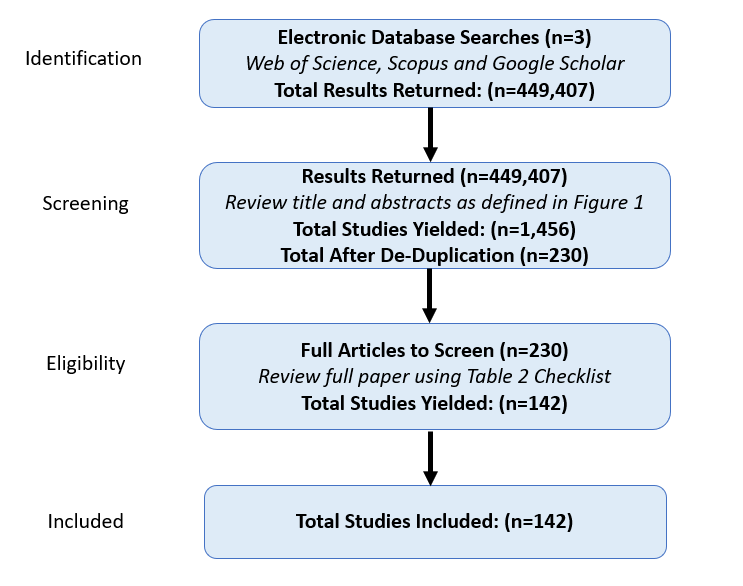}		\label{fig:Figure2}
\end{figure}

\section{Data}
\label{Data_3}
The classification of studies was independently conducted by the lead author (G.L.D.) who throughout the process conferred with all coauthors.
Each of the 142 studies in the set were reviewed and assigned labels to nine key features: {\bf Publication Year}, {\bf Category}, {\bf Context}, {\bf Demographics}, {\bf Target Region}, {\bf Data Source}, {\bf Theoretical Approach}, {\bf Methodologies}, and {\bf 4Ps}. We next explain each feature and its respective levels.\\

\noindent \underline {\bf Publication Year}: Observed years were \textbf{2010 -- 2021}, inclusive. Only studies available through December 2021 were included in our scope to allow for comparison across complete calendar years. This feature offers valuable information about the progress and patterns of research in OR and Analytics over more than ten years.\\

\noindent \underline {\bf Category}:
We considered three categories: \textbf{Operations Research (OR)}, \textbf{Analytics}, and \textbf{Position / Thought}. Whereas the first two categories are distinguished by their \textit{methodological} focus, the latter includes \textit{position / thought} pieces from either the OR or Analytics domains. We required a single category to be assigned to each study, and thus selected the category that we felt best matched the primary theme of the work.\\

\noindent\underline {\bf Context}:
We classified the primary topical HT area (as stated or inferred) into three contexts: {\bf Sex}, {\bf Labor}, and {\bf Both}. If the application was not specified, we assumed it to be general and thus applicable to both sex and labor trafficking. \\

\noindent\underline {\bf Demographics}: We classified studies into five demographic groups based on the population of interest (such as victims, potential victims, and survivors): {\bf Female}, {\bf Male}, {\bf Child}, {\bf LGBTQ+}, and {\bf Unspecified / All Individuals}. If no specific demographic characteristics were stated or could be inferred, we assumed it to be applicable to all individuals. A study could be classified into multiple demographics such as a study focused on female children. \\

\noindent\underline {\bf Target Region}: We subdivided the geographic location specified either by the data used in the study or by the region discussed in the background of the study, into world regions: {\bf Africa}, {\bf Asia}, {\bf Australia/Oceania}, {\bf Europe}, {\bf North America}, {\bf South America}, {\bf Unspecified / All regions}. A study could cover multiple geographic locations and therefore have multiple target regions identified. \\

\noindent\underline {\bf Data Source}: We classified the type of data used in the study into four categories: {\bf Primary}, {\bf Secondary}, {\bf Mixed}, {\bf N/A}. Primary data are collected directly from anti-trafficking organizations or researchers, including interviews and surveys. Secondary data are data that have already been collected for other purposes or are publicly available such as data from websites hosting illicit advertisements (such as backpage.com and rubmaps.com) and government reports. While many studies used their own methodologies to scrape public data sources such as escort and massage websites, we still consider these to be secondary sources. We classify studies utilizing expert judgements for determining data estimates to be secondary data. Mixed data means the study used both primary and secondary data in their work, and N/A indicates data was not used in the study.\\

\noindent\underline {\bf Theoretical Approach}: We classified the central theoretical approach to address HT into six categories: {\bf Decision Support}, {\bf Inferential Statistics / Detection}, {\bf Network Flow}, {\bf Resource Allocation}, {\bf Supply Chain}, and {\bf Other / Unspecified}. Decision Support explores ways to inform decision makers about initiatives to improve and better address HT, often building tools or systems for practitioners to use. Inferential Statistics / Detection focuses on identifying, estimating, or inferring aspects of HT. Network Flow studies are related to the flow of individuals and possibly trafficking network interaction. Resource Allocation addresses the use and allocation of resources in anti-HT efforts. Supply Chain studies examine the supply and demand of HT within a network. If a study approaches HT from a theoretical approach not listed, we label these works as Other / Unspecified. A study may be classified under multiple theoretical approaches. \\

\noindent\underline {\bf Methodologies}: We classified the main methodologies used in the set of studies into 21 categories: 

     \begin{multicols}{2}
    \begin{itemize}
        \item Active Learning
        \item Clustering or Classification
        \item Data Envelopment Analysis
        \item Empirical Analysis
        \item Facility Location
        \item Game Theory
        \item Graph Construction
        \item Investigative Search
        \item Link Inference
        \item Machine / Deep Learning (General)
        \item Natural Language Processing
        \item Information Extraction
        \item Integer Programming
        \item Network / Graph Theory
        \item Network Interdiction
        \item Queueing Theory
        \item (Social) Network Analysis
        \item Simulation
        \item Unsupervised or Minimally\\ \tabto{2em}Supervised Learning
        \item Web Crawling / Scraping
        \item Other
    \end{itemize}
    \end{multicols}

We ascribe methods to a study based on the introduction, conclusion, and main method or focus throughout the study. A study may apply a variety of different methods and therefore be classified into multiple methodologies.\\

\noindent\underline {\bf 4Ps}:
Activities to fight HT are often discussed under four broad principles: prevention, protection, prosecution, and partnership. These principles are collectively referred to as the 4Ps – a well-recognized classification within the anti-trafficking community~\citep{kon17}. A study may be classified under multiple principles. 
The 4Ps naturally correspond with efforts in the social science, healthcare, and law enforcement disciplines, and their alignment with OR and Analytics works is less evident. Thus, we adapt the 4Ps definitions to define each as it relates to OR and Analytics using the collective knowledge and experience of the authors in the anti-HT and OR and Analytics fields. To the best of our knowledge, this is the first attempt to define each of the 4Ps as it relates specifically to the OR and Analytics fields and constitutes an important contribution of this work.
Prevention, Protection, and Prosecution were originally referred to as the 3P paradigm~\citep{omc2016} which has since been informally expanded to include a fourth "P" representing Partnership. Prevention refers to efforts focused on a proactive approach to prevent trafficking such as awareness campaigns and education; Prosecution refers to efforts to punish traffickers; and Protection involves meeting post-trafficking victim needs such as counseling, job training, housing, and other support to facilitate survivor recovery and restoration.

Partnership was introduced to serve as a complementary means to further improve the efficacy among the 3Ps, enlisting all segments of society in the fight against HT~\citep{omc2016}. Together the 4Ps capture the spectrum of efforts in combating HT and therefore are an important feature for our literature review. Accordingly, we have adapted these 4Ps and classified studies in the following manner: 

\begin{itemize}
  \item {\bf Prevention:} The goal of the study is the prevention of HT either now or in the future and assumes no trafficking is currently taking place. Such studies typically feature victim-centric methodologies to help potential victims avoid being trafficked, such as awareness campaigns and education. Studies that consider reducing the re-trafficking risk of survivors who have left their trafficking environment also fall within the scope of the prevention principle.

  \item  {\bf Protection:} The goal of the study is to protect and aid the survivor during and post-exploitation. We consider victim-driven detection and disruption of HT networks to be a form of protection, as the focus of the study is mitigating the risk to an individual of further exploitation (including studies that consider NGOs, healthcare, and other non-law enforcement detection).
  
  \item {\bf Prosecution:} The goal of the study is to aid the prosecution of traffickers (often from a law enforcement perspective). We consider detection and disruption of HT networks aimed at locating, understanding, and stopping traffickers under the prosecution principle.
  
  \item {\bf Partnership:} The goal of the study is to show the benefit of collaboration and data sharing across different sectors, countries, or groups working together toward the common goal of addressing one or more areas of HT.

\end{itemize}

\section{Implications and Observations}\label{Implications_4}
Of the 142 studies in the set, the majority (73.9\%) were categorized as \emph{Analytics}, with 15.5\% classified as \emph{Operations Research} and 10.6\% as \emph{Position / Thought}. Fig \ref{fig:Figure3} depicts this breakdown. Fig \ref{fig:Figure4} provides summary statistics on each of the nine key features. All percentages are calculated in relation to the total number of studies in the set (142) unless stated otherwise. As some studies contained multiple methods or were identified to belong to multiple levels of a feature, a feature may not always sum to 100.0\%. 

\begin{figure}[!ht]
		\centering
	\caption{Category Classification for the Set of 142 Studies.}
	\vspace{0 in}
 \includegraphics[trim = 0 0 0 0, scale=1.4]{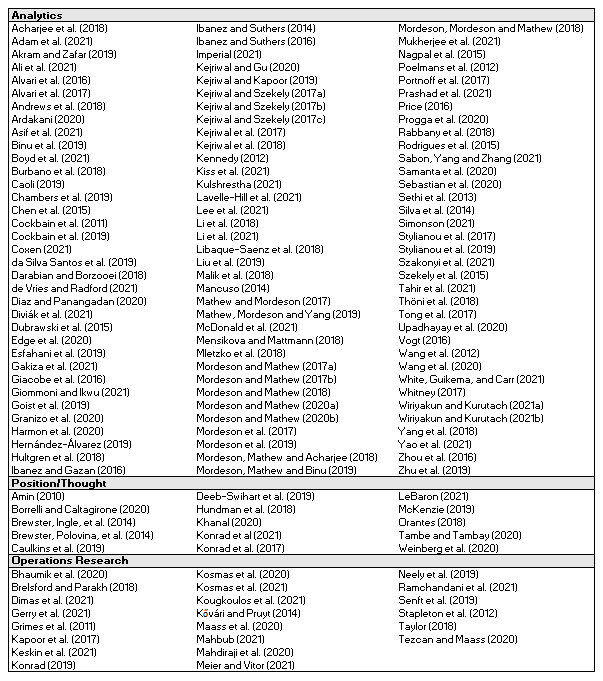}		\label{fig:Figure3}
\end{figure}

\begin{figure}[!ht]
		\centering
	\caption{Summary Statistics for the Set of 142 Studies.}
	\vspace{0 in}
 \includegraphics[trim = 0 0 0 0, scale=0.9]{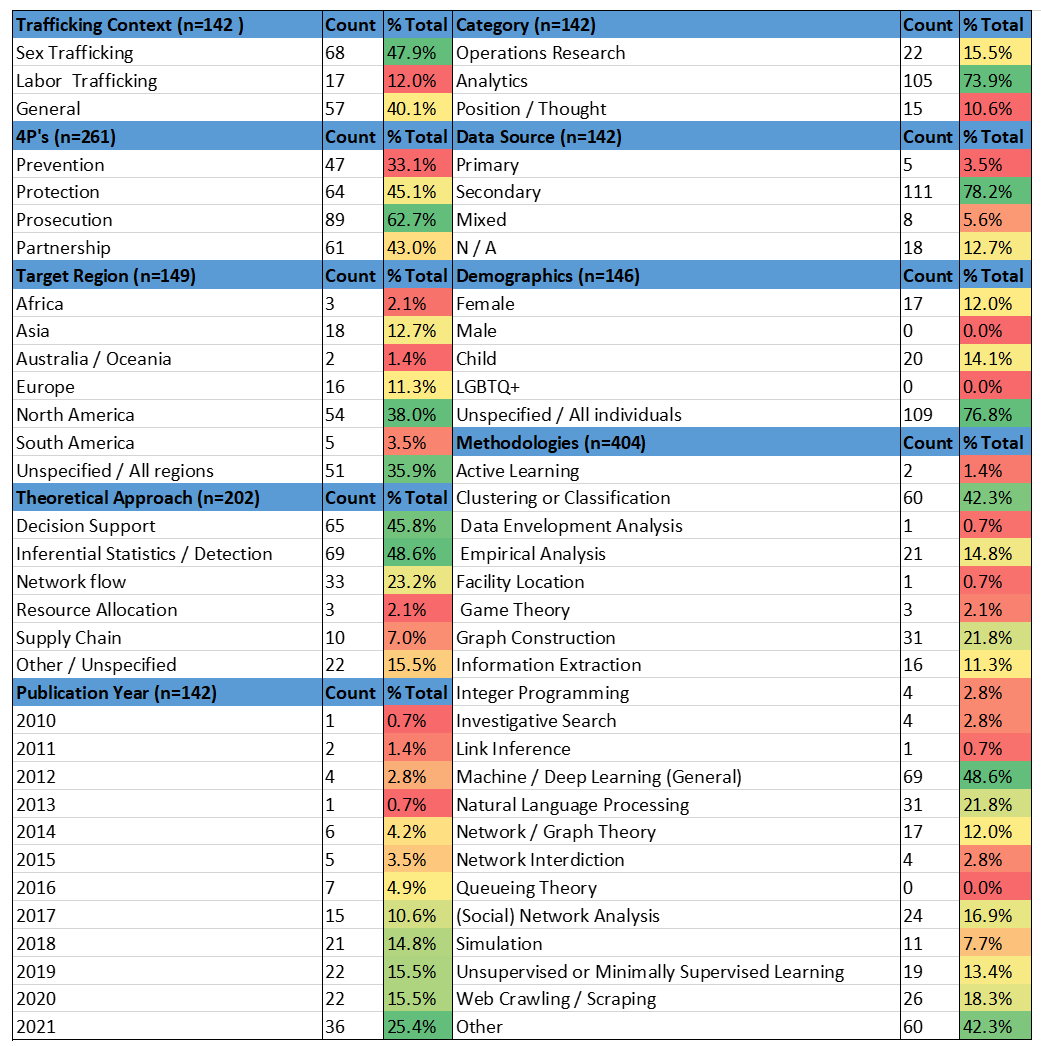}		\label{fig:Figure4}
\end{figure}

\subsection{Research Question 1: What Aspects of HT are Being Studied?}
\label{ResearchQuestion1_4_1}
        Although both sex and labor trafficking have been addressed in the OR and Analytics literature, an overwhelming number of studies focus specifically on \emph{sex trafficking}. Fig \ref{fig:Figure4} illustrates the inclination of OR and Analytics studies to focus on \emph{sex trafficking} (47.9\%), with only 12.0\% concentrating on \emph{labor trafficking}, while 40.1\% apply to both. As observed in Fig \ref{fig:Figure5}, studies overwhelming use \emph{secondary data}, with fewer than 4.0\% using a \emph{primary data} source. The use of \emph{secondary data} is likely due to accessibility; almost all studies on \emph{sex trafficking} (60.3\%) used data pulled from escort websites (or other online sites hosting illicit advertisements) which are public and therefore easier to access. The use of escort websites (in particular, backpage.com) as a source of data result in over 38.0\% of the studies focusing on the \emph{North American} region. Although the United States Department of Justice shut down backpage.com in 2018~\citep{usa18}, other escort and massaging sites offer illicit services and constitute the data source for several studies. Remarkably, 76.8\% of studies were not tailored to a specific demographic, despite the differences between typologies and demographics of victims \citep{pol17}. From the 4Ps perspective, \emph{prosecution} is the most common principle (62.7\%) among all studies, with considerably less focus on \emph{partnership}, \emph{protection}, and \emph{prevention} (Fig \ref{fig:Figure6}). A single study may be categorized under multiple  4Ps principles, and therefore the values in Fig \ref{fig:Figure6} do not sum to 100.0\%.

\begin{figure}[!ht]
		\centering
	\caption{Context of HT and Data Source.}
	\vspace{0 in}
 \includegraphics[trim = 0 0 0 0, scale=0.6]{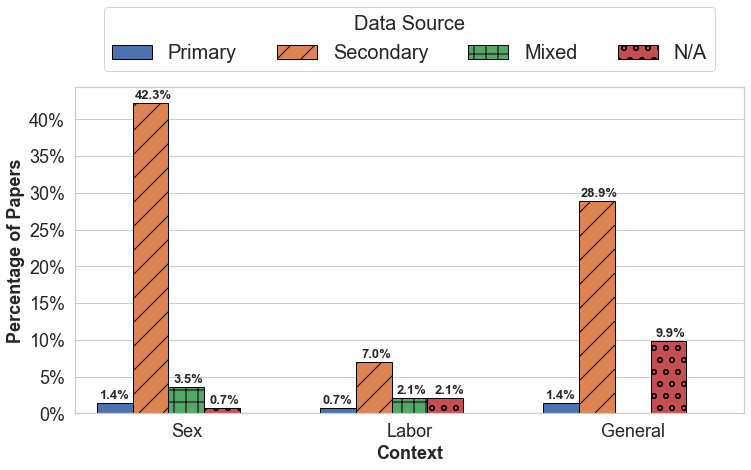}		\label{fig:Figure5}
\end{figure}

\begin{figure}[!ht]
		\centering
	\caption{Percent of Studies Involving 4Ps.}
	\vspace{0 in}
 \includegraphics[trim = 0 0 0 0, scale=0.5]{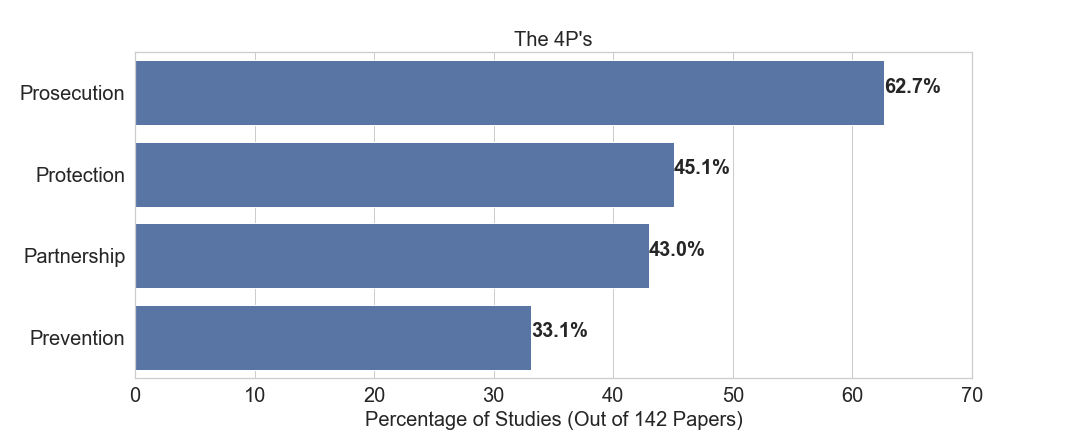}
 \label{fig:Figure6}
\end{figure}

\subsection{Research Question 2: What OR and Analytical Methods are Being Applied in the Anti-HT Domain?}
\label{ResearchQuestion2_4_2}

Fig \ref{fig:Figure4}  provides summary statistics such as counts and percentages for the frequency each category was observed in the set of studies. While the percentages are calculated based on the 142 studies in the set, because a study may belong to multiple classifications within a category, the total observations within each category are provided in the parenthesis. The colors highlight the magnitude of studies within each category, with green indicating higher percentages, and red lower percentages. A study may fall under more than one category and therefore the percentages will not always sum to 100.0\%. Methods related to machine learning (\emph{Machine / Deep Learning (General)}, \emph{Clustering or Classification}, \emph{Unsupervised / Minimally Supervised Learning}, \emph{Natural Language Processing}, and \emph{Active Learning}) were observed in over half of the studies. \emph{Machine / Deep Learning (General)} and \emph{Clustering or Classification} were the two most popular methods, accounting for about 32.0\% of all methods observed (out of the total 404 methods identified, see Fig \ref{fig:Figure4}). \emph{Web Crawling / Scraping} was used to generate a secondary dataset for analysis in 18.3\% of the studies, reflecting our previous observations in \ref{ResearchQuestion1_4_1} regarding the high use of \emph{secondary data} extracted from massage and escort websites. At least one method for the construction and analysis of networks (\emph{Graph Construction}, \emph{Network / Graph Theory}, \emph{Network Interdiction} and \emph{(Social) Network Analysis}) were observed in  34.5\% of the studies, with most emphasizing \emph{Graph Construction}.

The observed theoretical approaches are closely related to specific methods. For example, nearly half of the studies focused on \emph{Inferential Statistics / Detection} or \emph{Decision Support}, most of which applied various machine learning methods. \emph{Network Flow} methods appear in nearly 24.0\% of studies, specifically \emph{Graph Construction} and \emph{(Social) Network Analysis}.

Fig \ref{fig:Figure7} and Fig \ref{fig:Figure8} depicts each study on the $x$-axis. In Fig \ref{fig:Figure7} we display all studies categorized as \emph{Analytics}, and in Fig \ref{fig:Figure8} we categorize studies on the left as \emph{Operations Research} (in blue), and studies on the right as \emph{Position / Thought} studies (in red). For each study, the top six $y$-axis labels indicate its Theoretical Approach(es), while the remaining 21 following labels indicate methods used. If a study includes a given feature, the box is black, and grey otherwise. Theoretical approaches and methods are sorted in descending order based on the total count for each row.
Fig \ref{fig:Figure7} indicates that the majority of studies (over 77.0\%) in the \emph{Analytics} category take an \emph{Inferential Statistics / Detection} or \emph{Decision Support} theoretical approach. Fig \ref{fig:Figure8} shows that studies in the \emph{Operations Research} category are diverse, addressing the problem from distinct theoretical approaches and applying a variety of methodologies. Studies categorized as \emph{Position / Thought} address a variety of theoretical approaches and topics, which is a good indication that OR and Analytics researchers are exploring HT from different fields. 

While \emph{Resource Allocation} and \emph{Supply Chain} make up only around 9.0\% of all studies, they account for over 40.0\% within the \emph{Operations Research} category. Looking more closely at the relationship between \emph{Web Crawling / Scraping} and \emph{Clustering and Classification} methods we see a large majority (nearly 70.0\%) of \emph{Web Crawling / Scraping} studies apply \emph{Clustering and Classification} methods. In addition, more than 72.0\% of studies that applied both \emph{Web Crawling / Scraping} and \emph{Clustering and Classification} shared the goal of identifying \emph{sex trafficking} in online advertisements or tweets. 

 In order to compare studies across all three categories: \emph{Analytics}, \emph{Operations Research}, and \emph{Position / Thought} a combined figure of Fig \ref{fig:Figure7} and Fig \ref{fig:Figure8} can be found in the Appendix materials (see S1 Fig). We also provide a publicly available spreadsheet that can be used for closer examination of the individual studies, at {\url{https://github.com/gldimas/Dimas-et.-al-2022_Human-Trafficking-Literature-Review}}. This spreadsheet allows filtering studies on any combination of our nine features, returning all qualified studies. A screenshot of this tool can be see in Fig \ref{fig:Figure9}.

\newpage
\begin{landscape}
\centering
\begin{figure}
        \centering
    \caption{Granular View of Theoretical and Methodological Topic Inclusion for the Set of Studies Categorized as \emph{Analytics}.}
    
    \includegraphics[width=1.4\textwidth,height=0.95\textwidth]{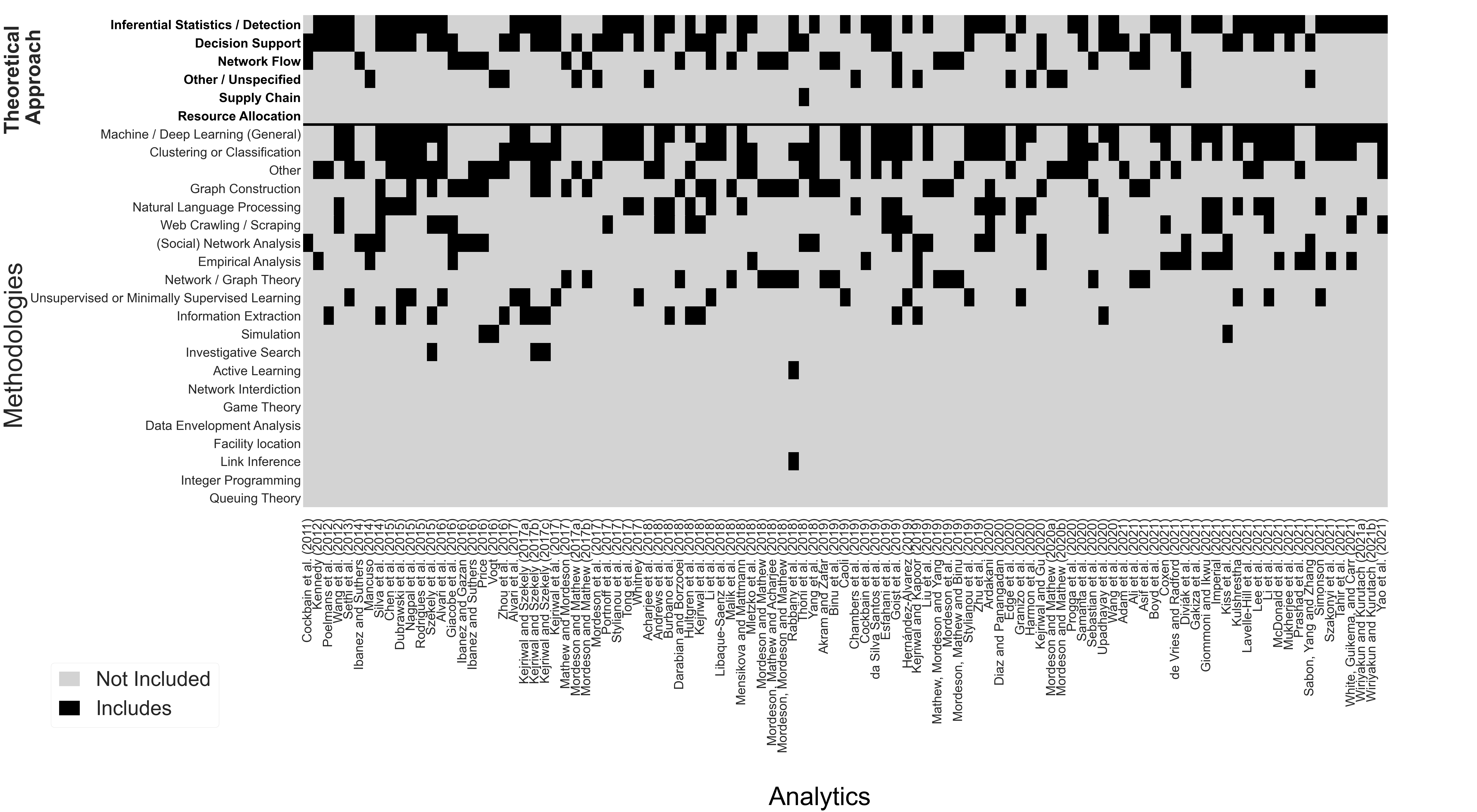}
    \label{fig:Figure7}
    {\scriptsize \emph{ The $x$-axis lists each paper;   and the $y$-axis depicts each of the Theoretical Approaches and Method. If a study includes a given feature, the box is black, and grey otherwise. Theoretical approaches and methods are sorted in descending order based on the total count for each row.}}
\end{figure}
\end{landscape}

\newpage
\begin{landscape}
\centering
\begin{figure}
        \centering
 \caption{\bf{Granular View of Theoretical and Methodological Topic Inclusion for the Set of Studies Categorized as \emph{Operations Research} or \emph{Position / Thought}}}
    
    \includegraphics[width=1.4\textwidth,height=0.95\textwidth]{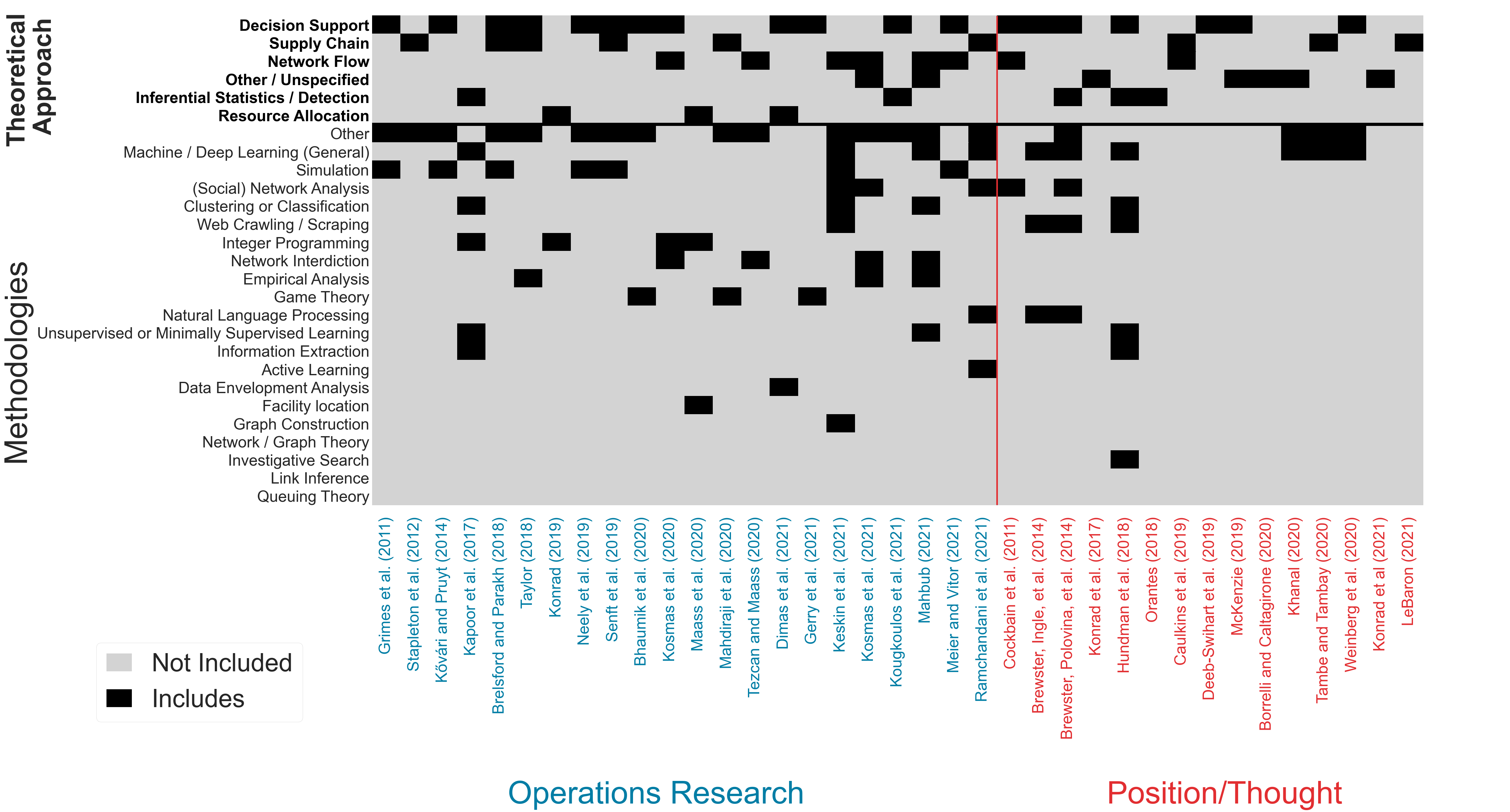}
    \label{fig:Figure8}
    {\scriptsize \emph{ The $x$-axis lists each paper;   and the $y$-axis depicts each of the Theoretical Approaches and Method. \emph{Operations Research} studies appear on the left (in blue), and \emph{Position / Thought} studies appear on the right (in red). If a study includes a given feature, the box is black, and grey otherwise. Theoretical approaches and methods are sorted in descending order based on the total count for each row.}}
\end{figure}
\end{landscape}

\newpage
\begin{landscape}
\centering
\begin{figure}[h!]
        \centering
  
    \includegraphics[width=1.40\textwidth,height=0.95\textwidth]{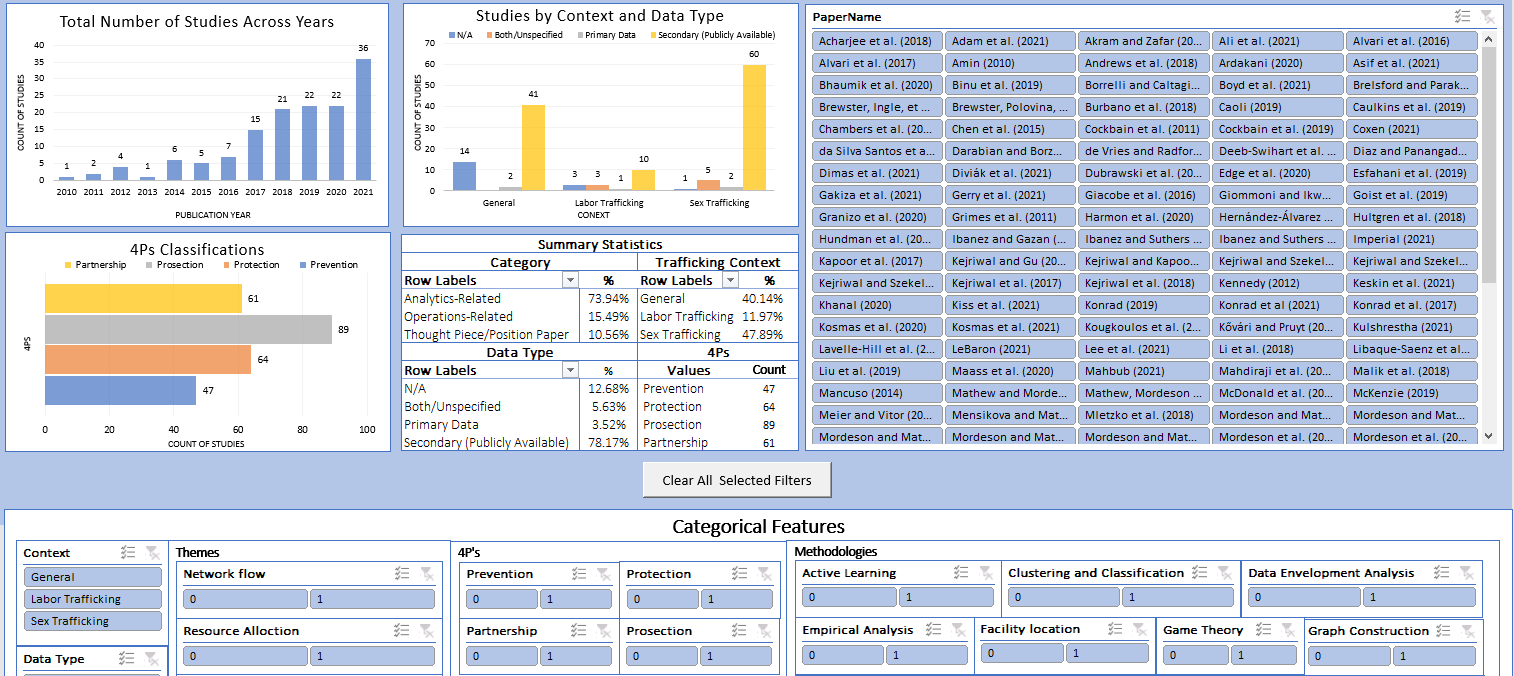}
     \caption{\bf{A Screenshot of The Spreadsheet Tool Created For Closer Examination of the Set of 142 Studies.}}
     \label{fig:Figure9}
    
\end{figure}
\end{landscape}

\subsection{Research Question 3: What are the Existing Gaps and Opportunities for Future Research?}
\label{ResearchQuestion3_4_3}

Several gaps emerged based on the classification and grouping of all 142 studies in the set on the prime principles (4Ps) (Fig \ref{fig:Figure6}), methods (Fig \ref{fig:Figure7}, Fig \ref{fig:Figure8}), trafficking context (Fig \ref{fig:Figure5}), and resulting observations to Research Questions 1 and 2. The following sections present opportunities for new avenues of investigation for OR and Analytics researchers in anti-HT efforts.  

\subsubsection*{Broaden the typology and demographics of trafficking studied }
\label{Broaden4_3_1}
The typology of trafficking activity, and by extension the demographic composition of victims, is diverse and under-explored in OR and Analytics~\citep{pol17}.
As noted in~\cite{kon21}, OR and Analytics researchers can increase the relevance and impact of their work by understanding the typology of trafficking and distinguish between various trafficking business models. To start, there is a clear need to expand the current focus to include labor trafficking. In OR and Analytics, sex trafficking (Fig \ref{fig:Figure6}) has received by far the most attention; while labor trafficking is estimated to account for over 60.0\% of all trafficking instances~\citep{int17}, it constitutes only 12.0\% of the reviewed studies. There is a close relationship between data source and type of HT studied. The absence of reliable, available data likely contributes to the lack of both analytically-based labor trafficking research, as well as the lack of diversity in data used in sex trafficking research; this is a well-documented issue across the HT literature \citep{swe18,far17,kon17,goz11,lac02}.  The implications of the lack of data are also evident in studies using OR and Analytics methods and discussed further in Section \ref{Diversify4_3_2}.

The diverse populations of those experiencing trafficking warrants further analysis. Nearly 77.0\% of the studies observed were classified as applicable to Unspecified / All individuals, indicating that the nuanced differences in victimology may be lacking. For example, no study we observed looked at trafficking through the lens of male victims or those who identify as LBGTQ+. These groups tend to be underrepresented in trafficking research despite their known presence~\citep{qui10,fre08}. Inclusivity of more diverse victims and trafficker demographics in trafficking research expands insights into the unique characteristics, needs and behaviors across trafficking.

Beyond the demographics of those impacted by trafficking, it is apparent that there is an opportunity to expand the diversity of the demographics of trafficking locations. Whereas trafficking occurs globally in various facets of society, and differs in its appearance across cultures, many of the sex trafficking studies we reviewed were conducted within developed countries (approximately 78.0\%). Thus, a clear opportunity exists to conduct anti-HT research in developing countries.

It is altogether possible that the lack of research identified in these areas may be a direct result of constrained factors such as the limited amount of data available. Even so, greater research diversity in trafficking typology and demographics will allow for an improved understanding of the extent and impact of trafficking worldwide, as well as greater insights into how future research can help address associated needs in the fight against HT. This brings us to the next research gap observed: the need to diversity data sources.

\subsubsection{Diversify the data sources}
\label{Diversify4_3_2}
Data collection and analysis have helped in the fight against HT in many ways, including the identification of HT victims \citep{hul18,zhu19,kes21}, informing prevention campaigns~\citep{kon19}, and detecting trafficking network behaviors \citep{coc11,li18}. While thorough data analysis lays the necessary groundwork for such discoveries, it relies upon the utilization of a variety of data from disparate sources. 

Over 78.0\% of all studies in the set used secondary data sources exclusively (Fig \ref{fig:Figure6}). The use of secondary data is common across many domains and proves beneficial given that the data already exists, is oftentimes publicly available, and can provide researchers with large amounts of data they might not be able to obtain otherwise. In the context of the OR and Analytics studies observed, over 22.0\% of the secondary data sources used were from the same source, the now defunct, backpage.com\citep{usa18}. Of the secondary data sources, around 54.0\% of these studies focused on sex trafficking. So, while secondary data sources have their place, as noted in the previous section the lack of data diversity compounds issues such as the typologies studied. Obtaining more robust data can broaden the information available, provide better insights into the scope of trafficking (such as prevalence), and examine changes that occur over time. Even with the limited current state of available data, OR and Analytics can still offer valuable contributions.

Researchers, particularly those in the Analytics community, could leverage collaborations and focus on developing the already existing, practitioner-collected data and help to operationalize it. For example, a common issue faced in HT analysis pertains to missing and incomplete data~\citep{ali10}. Many analytical methods exist that could help bridge this gap and improve the current anti-HT data landscape. Additionally, although many disparate datasets are available and growing, the anti-trafficking community rarely leverages combined data sources in a way to assess the status, trends, and dynamics of trafficking activities, and is another avenue for future work.  

Despite the usefulness of secondary data, they have drawbacks that may hinder the results of a study. Secondary data is often collected with another objective in mind and therefore the nuances of that data collection process may cause bias. For example, what constitutes trafficking may differ from study to study and is a well-documented issue across HT research (\citealt{agh08}, ~\citealt{alv12}, ~\citealt{gib20}). There may also be a context in which data simply does not exist. For these and other reasons, we advocate for the collection and use of primary data where possible. While primary data may be more resource-intensive to obtain, it provides researchers with curated data for their specific research goals. When possible, sharing this data within the anti-HT community provides additional resources for other work. In addition, embracing collaborations with practitioners or other researchers in the HT domain throughout the data collection phase can serve to increase the quality of this data by providing domain-specific insights. Building such collaborations demonstrates effective employment of the partnership principle (the fourth P) and the embodies our next recommendation.

\subsubsection{Better inclusion and collaboration of 4Ps }
\label{BetterInclusion4_3_3}

The 4Ps (prevention, protection, prosecution, and partnership) are widely acknowledged as a holistic set of principles that accounts for the spectrum of anti-HT efforts. To date, the majority of OR and Analytics studies in the set appear to be focused on prosecution (Section \ref{ResearchQuestion1_4_1}, Figure \ref{fig:Figure6}). Thus, while there exists demonstrated impact for prosecution-related activities, there are opportunities to contribute to anti-HT efforts in the spheres of prevention, protection, and partnership. A key way to increase the impact of OR and Analytics research in the fight against HT is to be keenly aware of all stakeholders involved, their various objectives, and how the research addresses the 4Ps. For example, while law enforcement may make decisions based on the likelihood of prosecuting traffickers, possibly at the expense of additional trauma to victims, Non-Government Organizations (NGOs) may focus more on the immediate needs of the survivors, offering an avenue for research around prevention and protection.

NGOs and governmental agencies often work directly with victims and survivors and could both inform avenues for profitable research studies, and themselves benefit from collaboration with OR and Analytics researchers. Given the often extreme resource constraints under which NGOs and governmental organizations operate, examining ways to evaluate current operations and improve resource allocation is a direction that deserves more study; less than 3.0\% of all studies considered these areas. 

Beyond the scope of the present work, OR and Analytics can help in the fight against HT by looking at push factors associated with HT. These areas include and are not limited to poverty, abuse, and lack of resources to meet basic needs~\citep{ker14}. More broadly, looking at ways to help improve circumstances of vulnerable populations at risk of HT is a needed avenue for future research.

While there are many ways in which the OR and Analytics communities can apply methods in the fight against HT, researchers ought to judiciously evaluate the problem context at hand, and whether an off-the-shelf method is justified; likely, the context warrants in-depth understanding, so that proper methodologies can be developed to accurately model and address the trafficking context~\citep{kon21}.

\section{Conclusions}
\label{Conclusions}
This survey provides a synopsis of the current state of the literature in OR and Analytics approaches in anti-HT contexts by surveying the research methodologies adopted in studies published from 2010 through 2021. A total of 142 studies were included in the set and examined, demonstrating the ability and promise of applying analytical methods to advance the fight against HT. A number of themes arose after careful review of the features of these studies, thereby illustrating opportunities for future research. We observed an increasing trend in publications for both OR and Analytics, thus demonstrating a growing awareness of the issue of HT. However, the tendency of these works to focus specifically on sex trafficking underscores the need for future research in labor trafficking. Very few (less than 24.0\%) of the studies on anti-HT in OR and Analytics focus on a specific sub-population, potentially failing to consider the diverse needs of victims and survivors. Existing OR and Analytics studies echo the anti-HT community at large for more available data. HT is diverse and nuanced, and researchers should make careful considerations when adapting existing methods to this vexing societal issue, considering efforts equally in prevention, protection, prosecution, and partnership.  

\ACKNOWLEDGMENT{We are grateful to the National Science Foundation (Operations Engineering grants CMMI-1841893 and CMMI-1935602) for their support. In addition, we express our deep gratitude for Dr. Meredith Dank at New York University for providing insights on the current research gaps in the domain of Human Trafficking and supplying feedback on how Operations and Analytical researchers can help fill these gaps.}


\bibliographystyle{pomsref} 
 \let\oldbibliography\thebibliography
 \renewcommand{\thebibliography}[1]{%
 	\oldbibliography{#1}%
 	\baselineskip14pt 
 	\setlength{\itemsep}{10pt}
 }
\bibliography{ref1.bib}



\nocite{ach18}\nocite{akr19}\nocite{ali21}\nocite{alv16}
\nocite{alv17}\nocite{ami10}\nocite{and18}\nocite{ard20}
\nocite{bha20}\nocite{bin19}\nocite{bor20}\nocite{boy21}
\nocite{bre18}\nocite{bre_Ing14}\nocite{bre_Pol14}\nocite{bur18}\nocite{cao19}\nocite{cau19}\nocite{cha19}\nocite{che15}
\nocite{coc19}\nocite{da19}\nocite{dar18}\nocite{dee19}
\nocite{dia20}\nocite{dub15}\nocite{edg20}\nocite{esf19}
\nocite{gal20}\nocite{ger21}\nocite{gia16}\nocite{goi19}
\nocite{gra20}\nocite{gri11}\nocite{har20}\nocite{her19}\nocite{hul18}\nocite{hun18}\nocite{iba_gaz16}\nocite{iba14}\nocite{iba_sut16}\nocite{kap17}\nocite{kej20}\nocite{kej19}\nocite{kej17a}\nocite{kej17b}\nocite{kej17c}\nocite{kej17}\nocite{kej18}\nocite{ken12}\nocite{ker14}\nocite{Dim21}\nocite{kha20}\nocite{kis20}
\nocite{kos20}\nocite{kou21}\nocite{kov14}\nocite{lav21}
\nocite{lib18}\nocite{liu19}\nocite{maa20}\nocite{mah20}
\nocite{mal18}\nocite{mat17}\nocite{mat19}\nocite{mcd21}
\nocite{mck19}\nocite{men18}\nocite{mle18}\nocite{mor17a}
\nocite{mor17b}\nocite{mor_mat18}\nocite{mor20a}\nocite{mor20b}
\nocite{mor_mat_mal17}\nocite{mor_sch19}\nocite{mor_mat_ach18}
\nocite{mor_mat_bin19}\nocite{mor_mor18}\nocite{muk21}
\nocite{nag15}\nocite{nee19}\nocite{ora18}\nocite{poe12}
\nocite{pra21}\nocite{pri16}\nocite{pro20}\nocite{rab18}
\nocite{ram21}\nocite{rod15}\nocite{sam20}\nocite{seb20}
\nocite{sen19}\nocite{set13}\nocite{sil14}\nocite{sim21}
\nocite{sta12}\nocite{sty17}\nocite{sty19}\nocite{sza21}
\nocite{sze15}\nocite{tam20}\nocite{tay18}\nocite{tez20}
\nocite{tho18}\nocite{ton17}\nocite{upa20}\nocite{vog16}\nocite{wan19}\nocite{wei20}\nocite{whi17}\nocite{yan18}\nocite{yao21}
\nocite{zho16}

\nocite{adam2021} \nocite{asif2021}\nocite{diviak2021}\nocite{gakiza2021}\nocite{giommoni2021}
\nocite{imperial2021}\nocite{coxen2021}\nocite{sabon2021}\nocite{kosmas2022}\nocite{lebaron2021}
\nocite{lee2021}\nocite{meier2021}\nocite{li2021}\nocite{de2022}\nocite{kulshrestha2021}\nocite{mahbub2021}
\nocite{white2021}\nocite{wiriyakun2021a}\nocite{wiriyakun2021b}\nocite{portnoff2017}\nocite{tahir2021}
\nocite{wang2012}


	
	

%
%
%
 \begin{APPENDICES}
{Appendix A: Supplementary Figure S1}
 \begin{landscape}
\centering

\begin{figure}
        \centering
    \caption{Granular View of Theoretical and Methodological Topic Inclusion for the Set of 142 Studies.}
    
    \includegraphics[width=1.40\textwidth,height=0.96\textwidth]{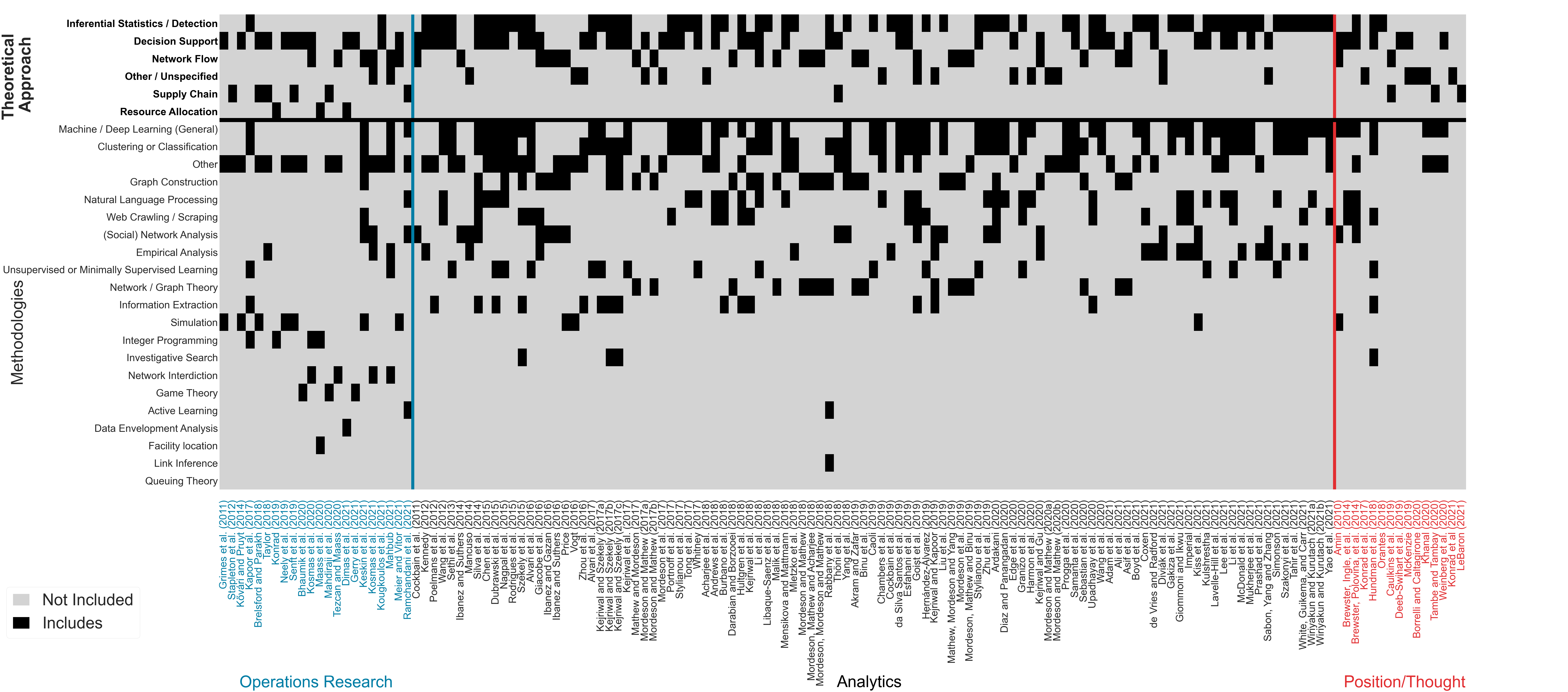}
    \label{fig:FigureS1}
    {\scriptsize \emph{This Figure depicts each study on the $x$-axis and each of the Theoretical Approaches and Method on the $y$-axis. \emph{Operations Research} studies appear on the far left (in blue), \emph{Analytics} studies appear in the middle (in black), and \emph{Position / Thought} studies appear on the far right (in red). If a study includes a given feature, the box is black, and grey otherwise. Theoretical approaches and methods are sorted in descending order based on the total count for each row.}}
\end{figure}
\end{landscape}
 \end{APPENDICES}

\end{document}